\documentclass[prd,preprint,a4paper,nofootinbib]{revtex4}

\usepackage{graphicx}
\usepackage{amsmath}
\usepackage{amsfonts}
\usepackage{amssymb}
\usepackage{bm}
\usepackage{bigstrut}
\def\ba{\begin{eqnarray}}
\def\ea{\end{eqnarray}}
\def\bq{\begin{equation}}
\def\eq{\end{equation}}

\def\sla#1{\ifmmode%
\setbox0=\hbox{$#1$}%
\setbox1=\hbox to\wd0{\hss$/$\hss}\else%
\setbox0=\hbox{#1}%
\setbox1=\hbox to\wd0{\hss/\hss}\fi%
#1\hskip-\wd0\box1 }

\begin{document}
\thispagestyle{empty}

\preprint{
\hbox to \hsize{
\hfill\vtop{\hbox{PITHA 04/08}
            \hbox{April 2004}} }
}

\title{\vspace*{.45in}
Top Background Extrapolation for $H\to WW$ Searches at the LHC}

\author{\vspace*{3mm}{\sc N.~Kauer}}
\affiliation{\vspace*{3mm}
Institut f\"{u}r Theoretische Physik E, RWTH Aachen, 52056
Aachen, Germany
\vspace*{.45in}}

\renewcommand{\abstractname}{{\normalsize \bf Abstract}}
\begin{abstract}
A leading order (LO) analysis is presented that demonstrates that key top backgrounds
to $H\to W^+W^-\to \ell^\pm \ell^\mp\sla{p}_T$ decays in weak boson
fusion (WBF) and gluon fusion (GF) at the CERN Large Hadron Collider
can be extrapolated from experimental
data with an accuracy of order 5\% to 10\%.  If LO scale variation
is accepted as proxy for the theoretical error, parton level results
indicate that the $t\bar{t}j$ background to the $H\to WW$ search
in WBF can be determined with a theoretical error of about
5\%, while the $t\bar{t}$ background to the $H\to WW$ search in 
GF can be determined with a theoretical error of better
than 1\%.  Uncertainties in the parton distribution functions
contribute an estimated 3\% to 10\% to the total error.
\end{abstract}

\maketitle


\section{Introduction \label{intro-section}}

Recent studies indicate that the CERN Large
Hadron Collider (LHC) will be able to discover
a Standard Model (SM) Higgs boson with mass between 100 and 200 GeV
with an integrated luminosity of only 10 to 30 fb$^{-1}$ if 
weak boson fusion (WBF) followed by $H\to\tau\tau$ and $H\to WW$
channels are taken into account \cite{WBFcits,KPRZ,LesHouchesHiggs2001,Asai2003}.
This intermediate mass range is currently
favored in light of a lower bound of 114.1 GeV from direct 
searches at LEP2 and an upper bound of 196 GeV from a SM
analysis of electroweak precision data (at 95\% CL)
\cite{lep2higgs}.  As discussed in detail in Ref.~\cite{LesHouchesHiggs2001},
Sec.~A.1, the precise knowledge of the
significance of any observed Higgs signal will require an
accurate determination of the SM backgrounds.
The $H\to WW\to \ell\ell\sla{p}_T$ decay channel ($\ell = e,\mu$)
in WBF as well as gluon fusion (GF) \cite{GFcits,DD2,ATLAS_TDR_2,LesHouchesHiggs2001}
is particularly challenging, because missing momentum prevents the
observation of a narrow mass peak that would allow an interpolation
of the backgrounds from side bands.
The purpose of this paper is to demonstrate how the extrapolation approach proposed in
Ref.~\cite{LesHouchesHiggs2001} can be applied to determine the
top quark background to the $H\to WW$ di-lepton decay mode at the LHC with an accuracy
that is in line with experimentalists' expectations.
In the remainder of this section we briefly describe our conventions and
the specifics of our calculations.  We can then quantify the theoretical
uncertainty of a conventional, leading order (LO) determination of the background
rates under consideration.  In Sections \ref{wbf-section} and \ref{gf-section},
we show for WBF and GF, respectively, how experimental data allows to determine 
these backgrounds with significantly reduced theoretical uncertainty.
In Section \ref{disc-section}, we consider caveats and improvements and conclude
with a summary in Section \ref{concl-section}.

To be specific, we consider the dominant $t\bar{t}$ + 1 jet background
to the $H\to W^+W^-\to \ell_1^\pm \ell_2^\mp\sla{p}_T$ search in WBF
and apply the selection cuts of Ref.~\cite{KPRZ} (see Sec.~\ref{wbf-section}),
which are very similar to the cuts adopted by the ATLAS and CMS collaborations.
We further consider the large $t\bar{t}$ background to the inclusive $H\to WW$ search,
i.e.~the same Higgs decay mode in GF.  In this case, we calculate results for
ATLAS selection cuts as given in Ref.~\cite{ATLAS_TDR_2},
Sec.~19.2.6, as well as CMS selection cuts \cite{DD2} (see Sec.~\ref{gf-section}).

To investigate the scale uncertainty of these backgrounds and how it
can be reduced we apply the following definitions for the
renormalization and factorization scales $\mu_R$ and $\mu_F$.
A factor $\xi$ is then used to vary the scales around the central values.
The suggestive scale choice for top production is the top mass $m_t = 175$ GeV:
\begin{equation}
\label{topscheme}
\mu_R = \mu_F = \xi m_t\,.
\end{equation}
Results for this scale choice are shown as solid curves in the figures.
For WBF, due to forward tagging selection cuts, the dominant
background arises from $t\bar{t}$ production with one additional hard jet.
To avoid double counting in this case, we alternatively calculate with
scales based on the minimal transverse mass:
\begin{equation}
\label{transscheme}
\mu_F = \xi \min(m_{T,t}, m_{T,\bar{t}}, p_{T,j}) \quad\text{and}\quad
\alpha_s^3 = \alpha_s(\xi m_{T,t})\alpha_s(\xi m_{T,\bar{t}})\alpha_s(\xi p_{T,j})\,.
\end{equation}
Results for this second definition are shown as dashed curves in the figures.
In principle, the renormalization and factorization scales are independent.  
We find, however, that the strongest scale variation occurs if both scales
are varied in the same direction and thus only introduce a single parameter $\xi$.
Scale-dependent quantities
are customarily condensed into the form $\hat{x} \pm\Delta \hat{x}$
based on a particular low and high scale choice.  We use the convention
\begin{equation}
\label{theoryerrorestimate}
\hat{x} = (x(\xi = \frac{1}{2})+x(\xi = 2))/2 \quad\text{and}\quad
\Delta \hat{x} = |x(\xi = \frac{1}{2})-x(\xi = 2)|/2\;,
\end{equation}
where $x$ is a cross section or cross section ratio.

All cross sections are calculated using the parton-level Monte Carlo
programs of Refs.~\cite{KZ1} and \cite{K2}, which include finite width effects and
the complete LO matrix elements for $\ell_1^\pm \ell_2^\mp\nu\bar{\nu}b\bar{b}$
(+ jets) final states.  We calculate with complete matrix elements unless otherwise
noted and use the complex mass scheme (CMS) \cite{cms-scheme} to
guarantee gauge invariance.\footnote{%
In Ref.~\protect\cite{K2}, we showed that the finite width scheme uncertainty,
i.e.~deviations due to different prescriptions to include finite width effects,
is smaller than 1\% for the backgrounds considered here.
We hence neglect it in this study.}
SM parameters and other calculational details are as described in Ref.~\cite{K2},
except that we use the updated parton distribution function (PDF) set
CTEQ6L1.\footnote{Note that CTEQ4L
was employed in Ref.~\protect\cite{KPRZ}.}  The calculations take into account
finite resolution and $b$ decay effects and a suboptimal $b$ tagging efficiency
$\varepsilon_{btag}$ based on expectations for the ATLAS and CMS detectors.

Figs.~\ref{wbf-scalevariation-40percent-btagging}(a),
\ref{wbf-scalevariation-60percent-btagging}(a),
\ref{gf-scalevariation-atlas-cuts}(a) and \ref{gf-scalevariation-cms-cuts}(a)
show the large scale variation that is expected for the LO 
background cross sections in both search channels.  For the WBF search channel,
the scale scheme (\ref{topscheme}) yields a background cross section of
$0.37 \pm 0.15$ fb, whereas the
scheme (\ref{transscheme}) yields
$0.57 \pm 0.25$ fb.  The theoretical uncertainty is 40--45\%.
Since the second cross section is not consistent with the first within 1$\sigma$,
it seems more appropriate to apply the prescription (\ref{theoryerrorestimate})
to the envelope of both curves.  All subsequent WBF results will be given using 
this procedure.  Then,
one obtains $0.52 \pm 0.30$ fb, with an even larger uncertainty of 60\%.
These results assume a $b$ tagging efficiency of 40\%.
With a more optimistic assumption of 60\% one obtains 
a 27\% smaller background with similar uncertainty: $0.38 \pm 0.22$ fb.
For the top background in the inclusive $H\to WW$ search
a somewhat smaller theoretical uncertainty is obtained, i.e.~3.4 fb (4.4 fb) with
an uncertainty of 25\% (25\%) for ATLAS (CMS) selection cuts
(with $\varepsilon_{btag}=50\%$).
For both channels it is obvious that the accuracy of
theoretical background calculations at LO is insufficient to determine the
total background with an accuracy of order 10\%, as required by experimental
physicists \cite{Asai2003}.


\section{Top background to $H\to WW$ decay in weak boson fusion \label{wbf-section}}

The extrapolation approach allows a more accurate determination
of a background cross section $\sigma_{bkg}$ if a
reference selection with a corresponding well-defined, measurable event 
rate $\sigma_{ref}\cdot{\cal L}$ can be found, so that the theoretical
uncertainty of the ratio $\sigma_{bkg}/\sigma_{ref}$ is small and a
sufficient number of events are observed during the run that
$\sigma_{ref}$ can be measured with low experimental uncertainty.\footnote{We neglect $\Delta{\cal L}$ and other systematic experimental uncertainties.}
The background cross section can then be approximated through
\begin{equation}
\label{extrapolationapproximation}
\sigma_{bkg} \quad \approx \quad \underbrace{\left(
     \frac{\sigma_{bkg,\text{ LO}}}{\sigma_{ref,\text{ LO}}}\right)}_{\stackrel{\mbox{\footnotesize low theoret.}}{\mbox{\footnotesize uncertainty}}}
     \quad\cdot\quad
\underbrace{\sigma_{ref}}_{\stackrel{\mbox{\footnotesize low experim.}}{\mbox{\footnotesize uncertainty}}}\;.
\end{equation}
Qualitatively, the smaller the difference between the cuts for
background and reference selection, the lower the uncertainty of
$\sigma_{bkg}/\sigma_{ref}$.
On the other hand, the selection cuts have to be modified sufficiently,
so that $\sigma_{ref}$ can be measured with good accuracy.
Thus, to derive suitable reference
selections from the corresponding background selections in the case at hand,
we propose the following strategy:
The WBF and inclusive $H\to WW$ search channel top
backgrounds are effectively suppressed through a central jet
veto.  Discarding this veto leads to a sizable increase of the
cross sections.  Secondly, to identify the top backgrounds in both cases,
we require that only events be considered that contain at least
one identified $b$ jet.   In our calculations we assume
that each $b$ (or $\bar{b}$) quark can be identified independently with probability $\varepsilon_{btag}$
if it is in the phase space region with $b$ tagging detector capability, which we
assume to be
\begin{gather}
\label{btagcoverage}
  p_{T,btag} > 15 \text{ GeV},\ \eta_{btag} < 2.5 \;.
\end{gather}
The probability $P_{btag}$ for a parton-level event to fulfill the $b$ tagging criterion is then given by
\bq
\label{atLeastOnebQuarkTaggedProbability}
P_{btag} = \begin{cases}
1-(1-\varepsilon_b)^2& \text{if $b$ and $\bar{b}$ quark fulfill (\protect\ref{btagcoverage})},\\
\varepsilon_b& \text{if either $b$ or $\bar{b}$ quark fulfill (\protect\ref{btagcoverage})},\\
0& \text{if neither $b$ nor $\bar{b}$ quark fulfill (\protect\ref{btagcoverage})},
\end{cases}
\eq
and the reference cross section is calculated by integrating $P_{btag}\,d\sigma_{ref}$.
Since events that are identified as top production via $b$ tagging can be eliminated 
from the signal sample, we calculate all background cross sections by
integrating $(1-P_{btag})\,d\sigma_{bkg}$.
If, after demanding a tagged $b$ jet and discarding the central jet veto, the
resulting reference rate is still too small, we also discard the lepton pair cuts.

\begin{figure}[htbp]
\begin{center}
\begin{minipage}[c]{.49\linewidth}
\flushright \includegraphics[width=6.cm, angle=90]{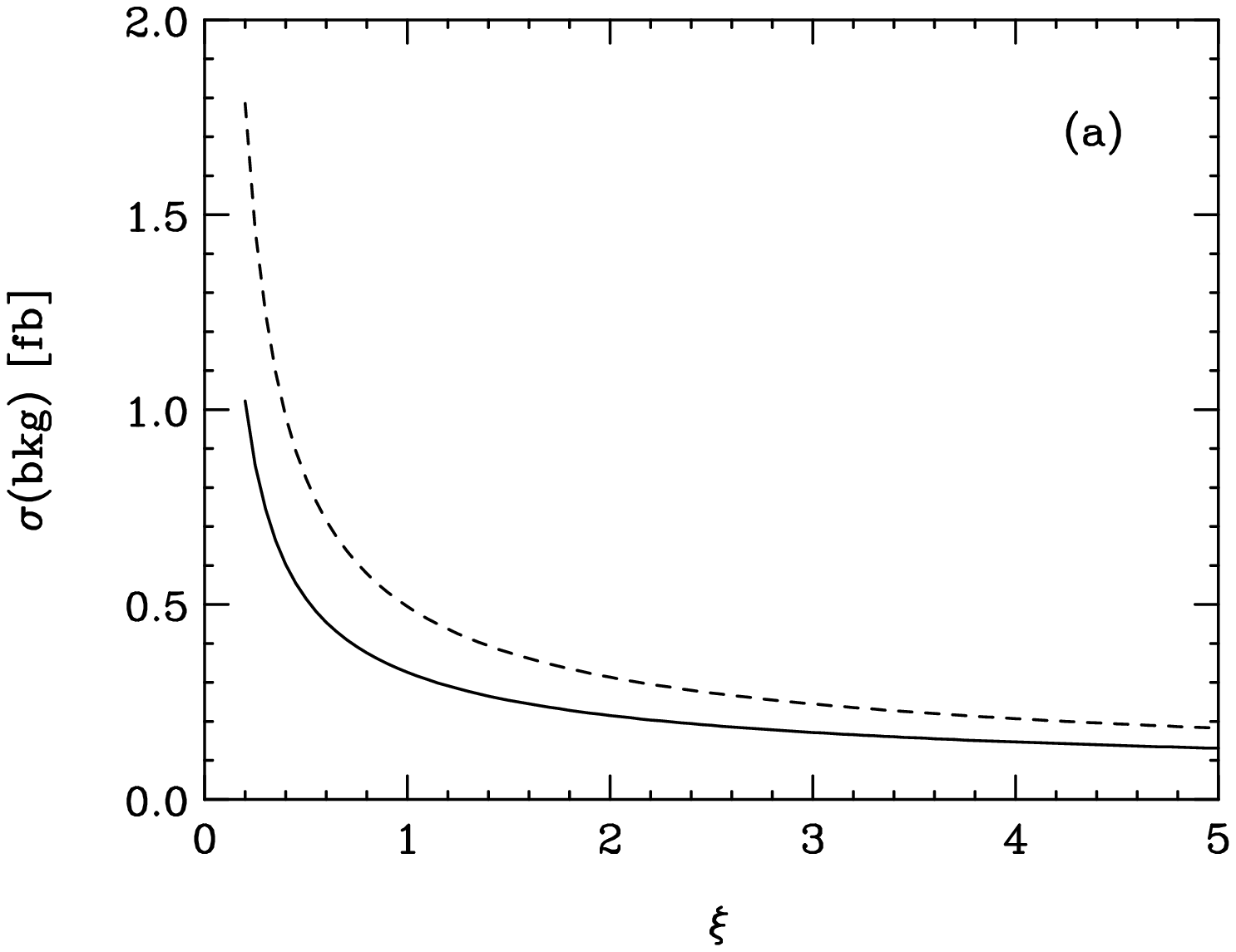}
\end{minipage} \hfill
\begin{minipage}[c]{.49\linewidth}
\flushleft \includegraphics[width=6.cm, angle=90]{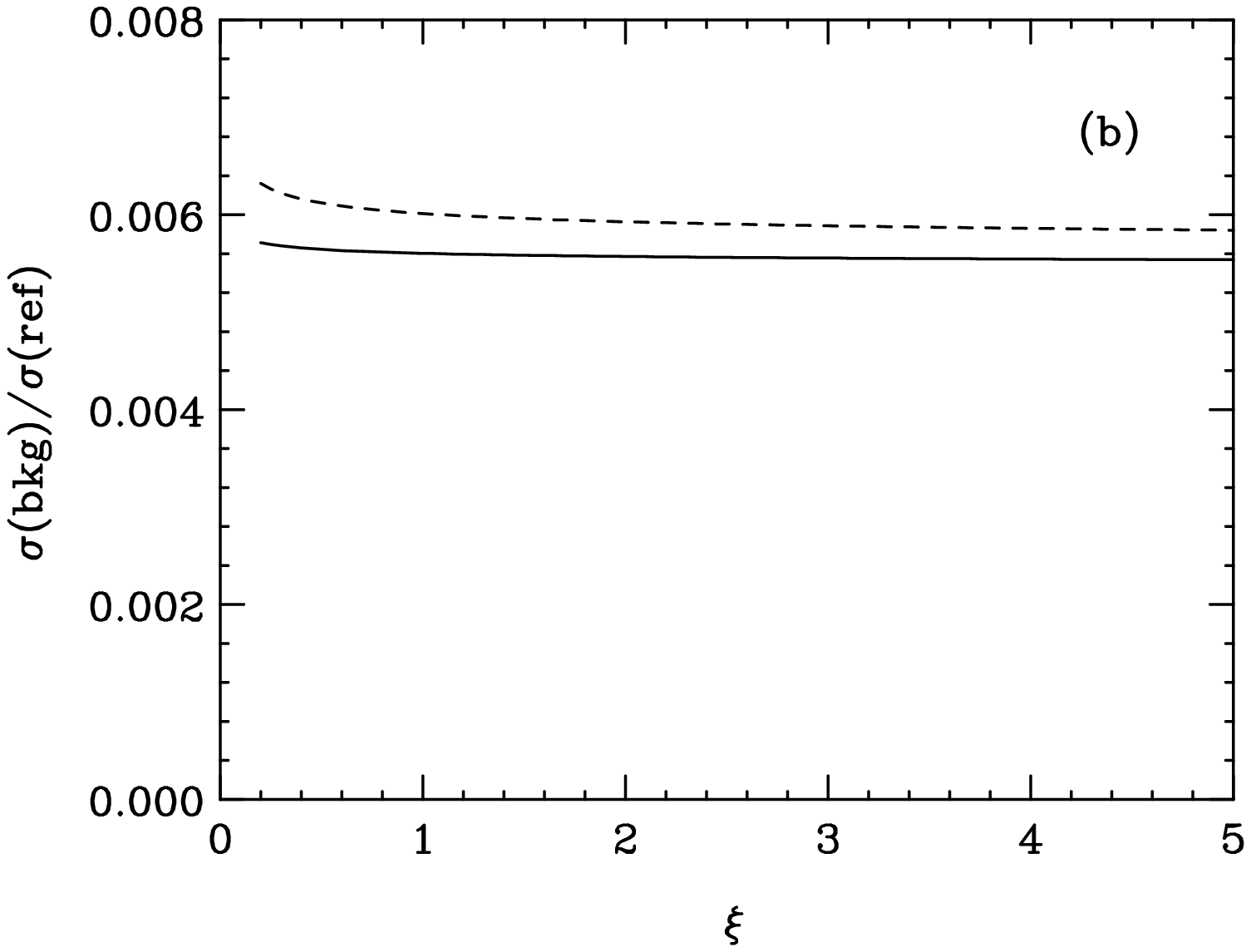} 
\end{minipage}
\caption{
  Renormalization and factorization scale variation of $t\bar{t}j$ background cross
  section (a) and ratio with reference cross section (b) to 
  $H\to W^+W^-\to \ell_1^\pm \ell_2^\mp\sla{p}_T$ search in weak boson fusion at the LHC
  for different scale definitions (see main text) and $\varepsilon_{btag} = 40\%$.
}
\label{wbf-scalevariation-40percent-btagging}
\end{center}
\end{figure}

\begin{figure}[htbp]
\begin{center}
\begin{minipage}[c]{.49\linewidth}
\flushright \includegraphics[width=6.cm, angle=90]{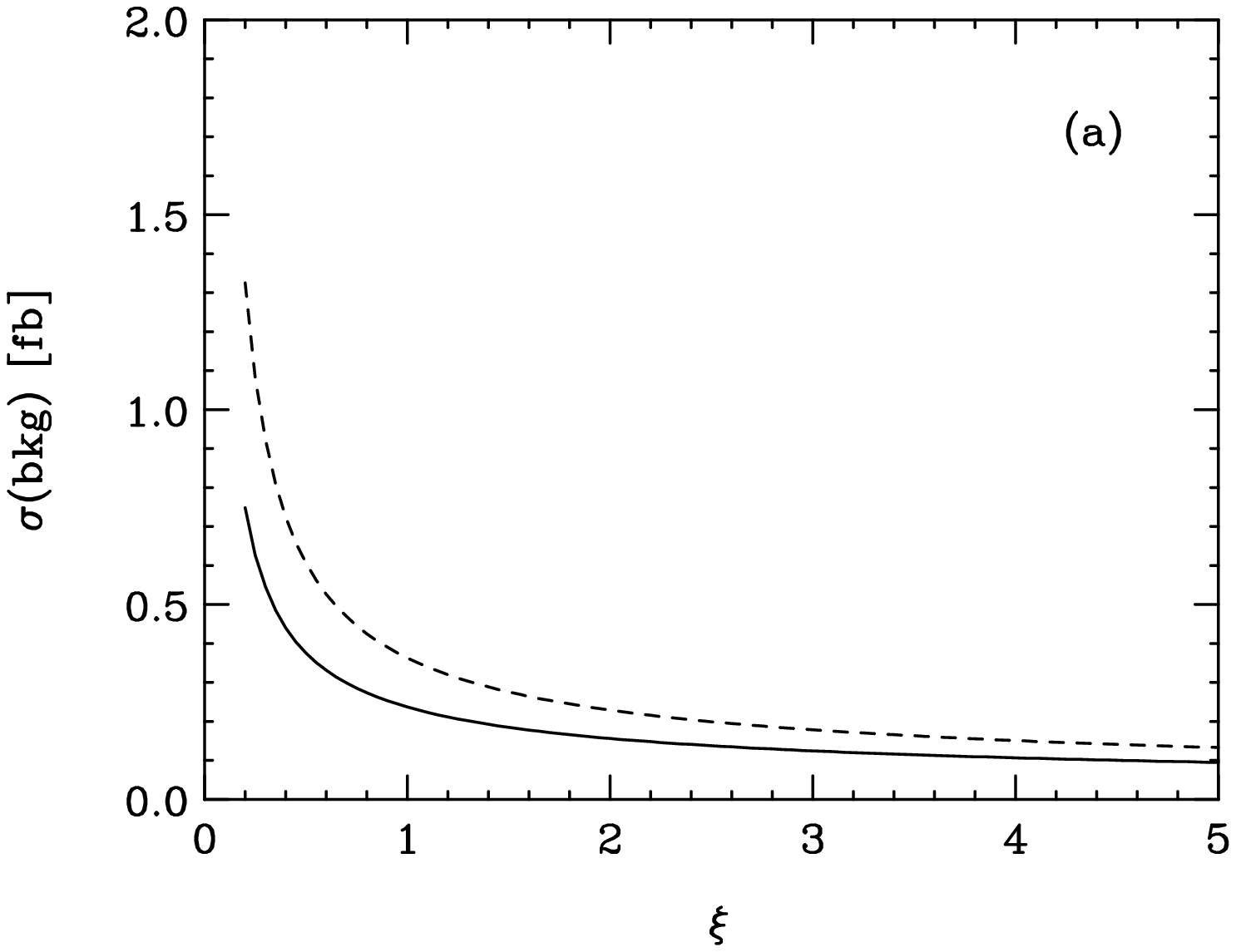}
\end{minipage} \hfill
\begin{minipage}[c]{.49\linewidth}
\flushleft \includegraphics[width=6.cm, angle=90]{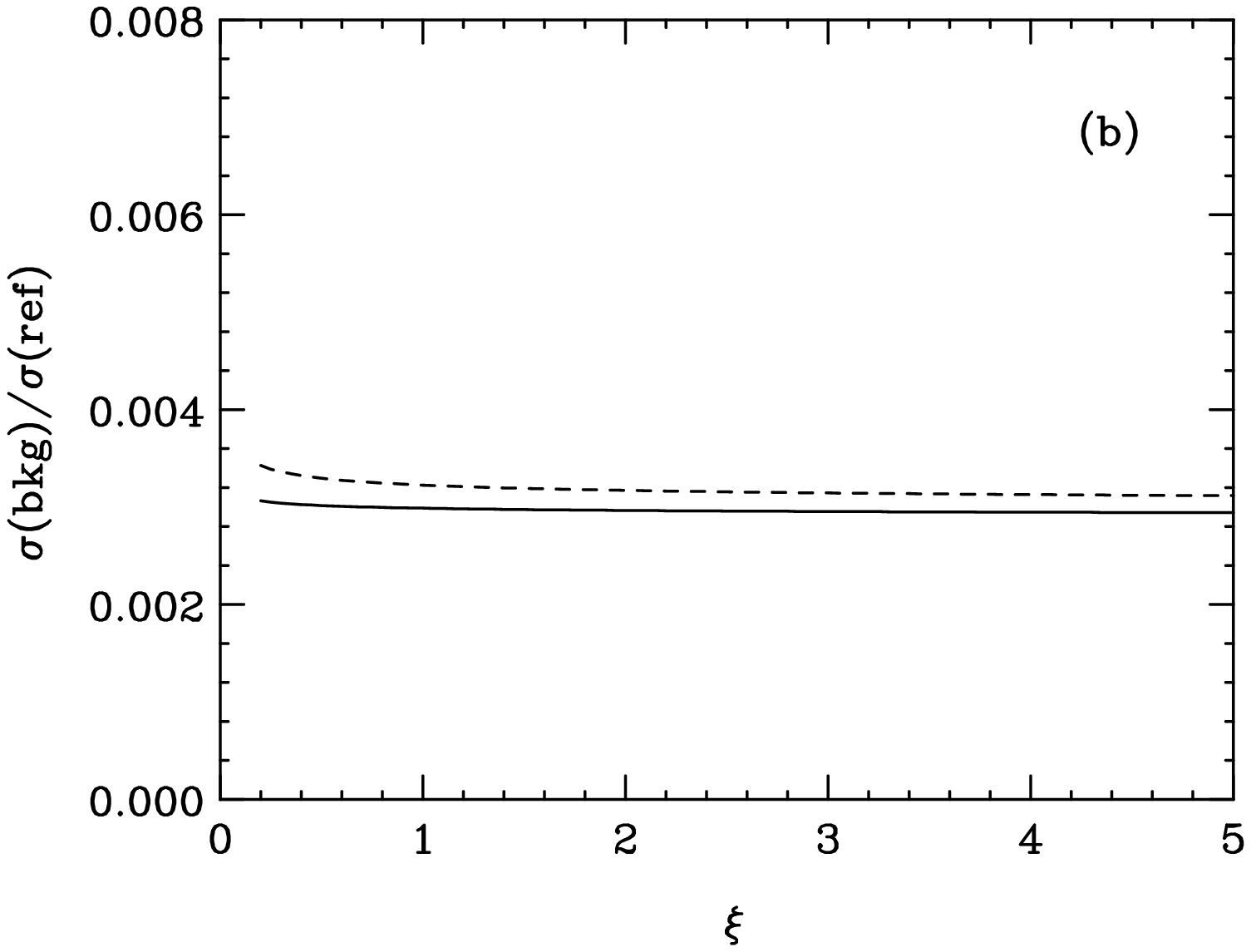} 
\end{minipage}
\caption{
As Fig.~\protect\ref{wbf-scalevariation-40percent-btagging}, but for
$\varepsilon_{btag} = 60\%$.
}
\label{wbf-scalevariation-60percent-btagging}
\end{center}
\end{figure}

In the search for a light Higgs boson in WBF the selection is given by
the forward tagging cuts
\begin{gather}
p_{Tj} > 20 \text{ GeV},\ |\eta_j| < 4.5,\ \Delta R_{jj} > 0.6,\notag\\
p_{T\ell_1} > 20 \text{ GeV},\ p_{T\ell_2} > 10 \text{ GeV},\ |\eta_\ell| < 2.5,\ \Delta R_{j\ell} > 1.7,\notag\\
\eta_{j,min} + 0.6 < \eta_{\ell_{1,2}} < \eta_{j,max} - 0.6,\notag\\
\eta_{j_1}\cdot \eta_{j_2} < 0,\notag\\
m_{jj} > 600 \text{ GeV},\ |\eta_{j_1} - \eta_{j_2}| > 4.2,\notag\\
\sla{p}_T > 20 \text{ GeV }\quad\text{ provided } p_{TH} < 50 \text{ GeV}
\end{gather}
and the lepton pair cuts
\begin{gather}
  m_{\ell\ell} < 60 \text{ GeV},\ \Delta\phi_{\ell\ell} < 140^\circ,\notag\\
  x_{\tau_1} > 0,\ x_{\tau_2} > 0,\ m_{\tau\tau} > m_Z - 25 \text{ GeV},\notag\\
  50 \text{ GeV} < m_{T,1}(WW) < m_H + 20 \text{ GeV},\notag\\
  \Delta\phi(\ell\ell,\sla{p}_T) + 1.5\,p_{TH} > 180,\ \ 12\,\Delta\phi(\ell\ell,\sla{p}_T)+ p_{TH} > 360
  \label{wbf-lepton-pair-cuts}
\end{gather}
with $m_{T,1}(WW) := [(E_{T,\ell\ell}+\sla{E}_T)^2 - (\vec{p}_{T,\ell\ell} + \sla{\vec{p}}_T)^2]^{1/2}$
with transverse energies $E_{T,\ell\ell} = (p_{T,\ell\ell}^2 + m_{\ell\ell}^2)^{1/2}$ and
$\sla{E}_T = (\sla{p}_T^2 + m_{\ell\ell}^2)^{1/2}$.
We fix $m_H = 120$ GeV, which defines the transverse mass window cut.
The jet veto is applied by discarding all events where an additional jet is
located between the tagging jets,
\begin{gather}
\label{wbf-jet-veto}
  p_{Tv} > 20 \text{ GeV},\ \eta_{j,min} < \eta_v < \eta_{j,max} \;.
\end{gather}
The reference selection obtained by eliminating the veto (\ref{wbf-jet-veto})
and requiring at least one tagged $b$ jet yields a cross section of
13 fb, which, with 30 fb$^{-1}$, would result in a statistical
uncertainty for the measured rate of about 5\% (using Poisson
statistics).  We therefore also discard the lepton pair cuts
(\ref{wbf-lepton-pair-cuts}).
The resulting reference cross section of $118 \pm 66$ fb gives rise to
a statistical error of slightly less than 2\% with 30 fb$^{-1}$ of data
for $\varepsilon_{btag} = 60\%$ (and also with $87 \pm 48$ fb for $\varepsilon_{btag} = 40\%$).
Note that the scale uncertainty of these reference cross
sections is very similar to that of the background cross sections.
However, the scale variation of the corresponding ratios
$\sigma_{bkg}/\sigma_{ref}$ is significantly reduced as shown in
Figs.~\ref{wbf-scalevariation-40percent-btagging}(b) and \ref{wbf-scalevariation-60percent-btagging}(b).  One obtains $0.0059 \pm 0.0003$ for $\varepsilon_{btag} = 40\%$
and $0.0031 \pm 0.0002$ for
$\varepsilon_{btag} = 60\%$,
or a relative error of 5\%.  Note that the applicable ratio depends
strongly on the achieved $b$ tagging efficiency.\footnote{%
The details of $b$-tagged event rejection for $\sigma_{bkg}$ also strongly affect
the ratio.  If, for example, only events with $b$-tagged forward tagging jets are
discarded, background and ratio increase by 30\%.  The sensitivity to
variations in the gluon PDF is smaller: Calculating with CTEQ4L instead of CTEQ6L1
reduces the ratio by 7\%.}

As seen in Table \ref{wbf-measurement-uncertainty}, an integrated luminosity of 
30 fb$^{-1}$ would allow a measurement of the WBF reference cross section
with a statistical error of 2\% or less.
Combining the uncertainty of both extrapolation factors in quadrature
yields a WBF background estimate with an accuracy of about 5\%.

\begin{table}
\caption{Expected number of events $E$ and statistical experimental error for
WBF reference selection for different integrated luminosities.
\label{wbf-measurement-uncertainty}}
\vspace*{.5cm}
\begin{tabular}{|l|c|c||c|c|}
 \cline{2-5}
\multicolumn{1}{c|}{} & \multicolumn{2}{c||}{$\varepsilon_{btag} = 40\%$} &
\multicolumn{2}{c|}{$\varepsilon_{btag} = 60\%$} \\
 \hline
 $\int{\cal L}\;dt$ & $E$ & $\Delta E/E$ & $E$ & $\Delta E/E$ \\
 \hline
  10 fb$^{-1}$ & 870 & $\pm$ 3.4\% & 1180 &  $\pm$ 2.9\% \\
  30 fb$^{-1}$ & 2610 & $\pm$ 2.0\% & 3540 &  $\pm$ 1.7\% \\
  100 fb$^{-1}$ & 8700 & $\pm$ 1.1\% & 11800 &  $\pm$ 0.9\% \\
  \hline 
\end{tabular}
\end{table}

Since top backgrounds are often calculated with top quark in narrow width
approximation (NWA), we show in Table \ref{wbf-cms-nwa-comparison} the change
of background cross section and ratio for the WBF selection if sub- and
non-resonant amplitude contributions are omitted.  The complete off-shell
matrix element increase of 15\% for the background is reduced to 5\% for the
ratio---a level also found for inclusive top pair production at the LHC
(see Ref.~\cite{K2}).

\begin{table}
\caption{%
Change of background cross section and ratio for WBF selection using scale scheme
(\protect\ref{transscheme}) (with $\xi = 1$) if calculated with complete
tree-level matrix elements \protect\cite{KZ1,K2} relative to calculating with
top quark in narrow width approximation (NWA).
\label{wbf-cms-nwa-comparison}}
\vspace*{.2cm}
\begin{tabular}{|c|c|c|}
 \cline{2-3}
\multicolumn{1}{c|}{} & \multicolumn{2}{c|}{$x_\text{CMS}/x_\text{NWA}$ factor} \\
\hline
$x$  & $\varepsilon_{btag} = 40\%$ & $\varepsilon_{btag} = 60\%$ \\
 \hline
 $\sigma_{bkg}$ & 1.15 & 1.16 \\
 $\sigma_{bkg}/\sigma_{ref}$ & 1.05 & 1.05 \\
 \hline
\end{tabular}\\
\vspace*{.5cm}
\end{table}

Besides scale variation, a second, smaller source of uncertainty in the
background determination arises due to uncertainties in the PDFs.  
Since top pair production at the LHC is dominated by gluon
scattering, the large gluon density uncertainty for $x \gtrsim 0.2$ leads to
large uncertainties for theoretical cross section calculations.
In Ref.~\cite{K2}, we observed that PDF improvements lead to relative changes
of inclusive cross sections by 10-20\%, while cross section ratios of the type
shown in Table \ref{wbf-cms-nwa-comparison} are almost constant.  To properly
quantify the PDF uncertainty of an observable, an eigenvector basis approach
to the Hessian method can be used (see e.g.~Ref.~\cite{cteq6}).  Unfortunately,
this method is currently only available for NLO PDF sets, whereas a LO fit
is most appropriate for the calculations performed here. Nevertheless, to
provide an estimate for the PDF uncertainty of
$\sigma_{bkg}$ and $\sigma_{bkg}/\sigma_{ref}$, we 
show results in Table \ref{wbf-pdf-uncertainty} that use
the ``best fit'' PDF sets CTEQ6L1 (LO) and CTEQ6.1M (NLO), and in addition
use the corresponding eigenvector basis CTEQ61.01-40 to calculate PDF uncertainties
according to (3) in Ref.~\cite{cteq6}.  When comparing the results for
$\sigma_{bkg}$ and $\sigma_{bkg}/\sigma_{ref}$, one finds that the
relative error decreases from 12\% to about 5\%, while the relative deviation of
LO and NLO PDF results increases to about 10\%.  We therefore estimate the
PDF uncertainty of the WBF ratio $\sigma_{bkg}/\sigma_{ref}$ at 5-10\%.

\begin{table}
\caption{\label{wbf-pdf-uncertainty}
WBF top background cross section $\sigma :=
\sigma_{bkg}$ and cross section ratio $K:= \sigma_{bkg}/\sigma_{ref}$
calculated with PDF sets CTEQ6L1 and CTEQ6.1M (= CTEQ61.00)
using scale scheme (\protect\ref{transscheme}) (with $\xi = 1$).
The NLO sets CTEQ61.01-40 allow to calculate a PDF uncertainty for observables
(see main text).
}
\vspace*{.5cm}
\begin{tabular}{|l|c|c|c|c|}
\hline
\multicolumn{1}{|c|}{$\varepsilon_{btag} = 40\%$} & $\sigma$ &
$\displaystyle \frac{\Delta \sigma}{\sigma}$ & $K$ &
$\displaystyle \frac{\Delta K}{K}$ \bigstrut \\
 \hline
 CTEQ6L1 (LO) & 0.50 fb & -- & 0.0060 & -- \\
 CTEQ6.1M  (NLO) & 0.49 fb & $\pm$\:12\% & 0.0066 & $\pm$\:4.7\% \\
  \hline 
\end{tabular}\\[.5cm]
\begin{tabular}{|l|c|c|c|c|}
\hline
\multicolumn{1}{|c|}{$\varepsilon_{btag} = 60\%$} & $\sigma$ &
$\displaystyle \frac{\Delta \sigma}{\sigma}$ & $K$ &
$\displaystyle \frac{\Delta K}{K}$ \bigstrut \\
 \hline
 CTEQ6L1 (LO) & 0.36 fb & -- & 0.0032 & -- \\
 CTEQ6.1M  (NLO) & 0.37 fb & $\pm$\:13\% & 0.0036 & $\pm$\:6.4\% \\
  \hline 
\end{tabular}\\[.5cm]
\end{table}


\section{Top background to $H\to WW$ decay in gluon fusion \label{gf-section}}

The analysis of the extrapolation of the top background to the
$H\to WW$ di-lepton decay mode in gluon fusion proceeds along the
same lines as Sec.~\ref{wbf-section}.  For this Higgs search channel,
which is important for Higgs masses between 140 and 180 GeV, we consider
the selection cuts adopted by the ATLAS collaboration:
\begin{gather}
p_{T\ell_1} > 20 \text{ GeV},\ p_{T\ell_2} > 10 \text{ GeV},\ |\eta_\ell| < 2.5,\ 
\sla{p}_T>40 \text{ GeV},\notag\\
m_{\ell\ell}<80 \text{ GeV},\ \Delta\phi_{\ell\ell} < 1.0\ \text{rad},\ |\theta_{\ell\ell}| < 0.9\ \text{rad},\ |\eta_{\ell_1} - \eta_{\ell_2}| < 1.5, \notag\\
m_H - 30 \text{ GeV} < m_{T,2}(WW) < m_H
\label{gf-atlas-cuts}
\end{gather}
with
$m_{T,2}(WW) := [2p_T^{\ell\ell}\sla{p}_T(1-\cos\Delta\phi(\ell\ell,\sla{p}_T))]^{1/2}$
and the transverse mass window cut fixed by choosing $m_H = 170$ GeV in our
ATLAS calculations.  The ATLAS selection cuts also include a central jet veto,
that discards all events with jets that fulfill
\begin{gather}
\label{gf-atlas-jet-veto}
p_{Tv} > 15 \text{ GeV},\  |\eta_v| < 3.2 \;.
\end{gather}
We also present results for the selection cuts adopted by the CMS collaboration:
\begin{gather}
p_{T\ell_1} > 25 \text{ GeV},\ p_{T\ell_2} > 10 \text{ GeV},\ |\eta_\ell| < 2.4,\notag\\
\theta_{\ell\ell}>30^\circ,\ |\eta_{\ell_1} - \eta_{\ell_2}| < 1.25,\ \Delta\phi_{\ell\ell} < 45^\circ \;.
\label{gf-cms-cuts}
\end{gather}
Here, all events are discarded that have jets that fulfill
\begin{gather}
\label{gf-cms-jet-veto}
p_{Tv} > 20 \text{ GeV},\ |\eta_v| < 3 \;.
\end{gather}

\begin{figure}[htbp]
\begin{center}
\begin{minipage}[c]{.49\linewidth}
\flushright \includegraphics[width=6.cm, angle=90]{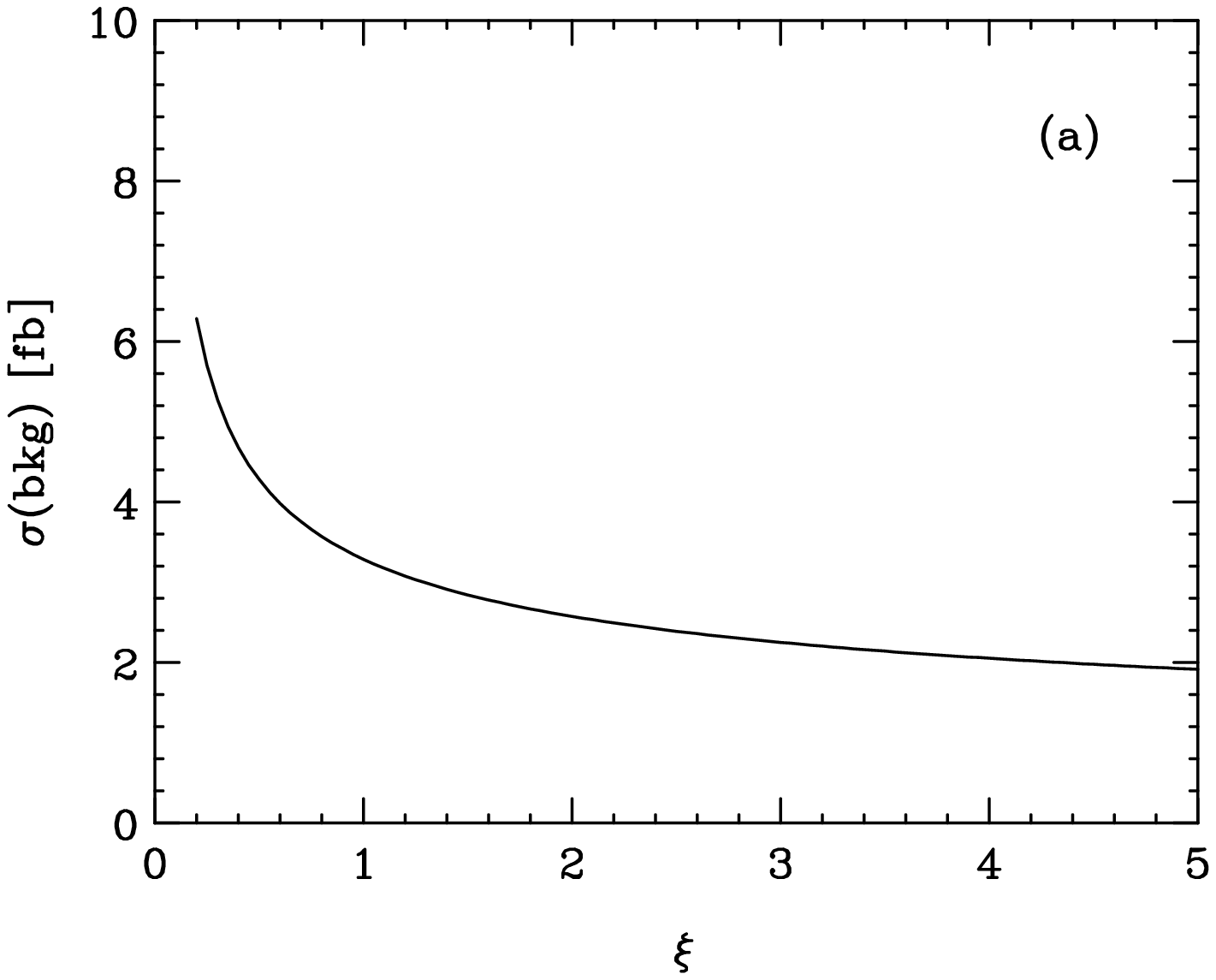}
\end{minipage} \hfill
\begin{minipage}[c]{.49\linewidth}
\flushleft \includegraphics[width=6.cm, angle=90]{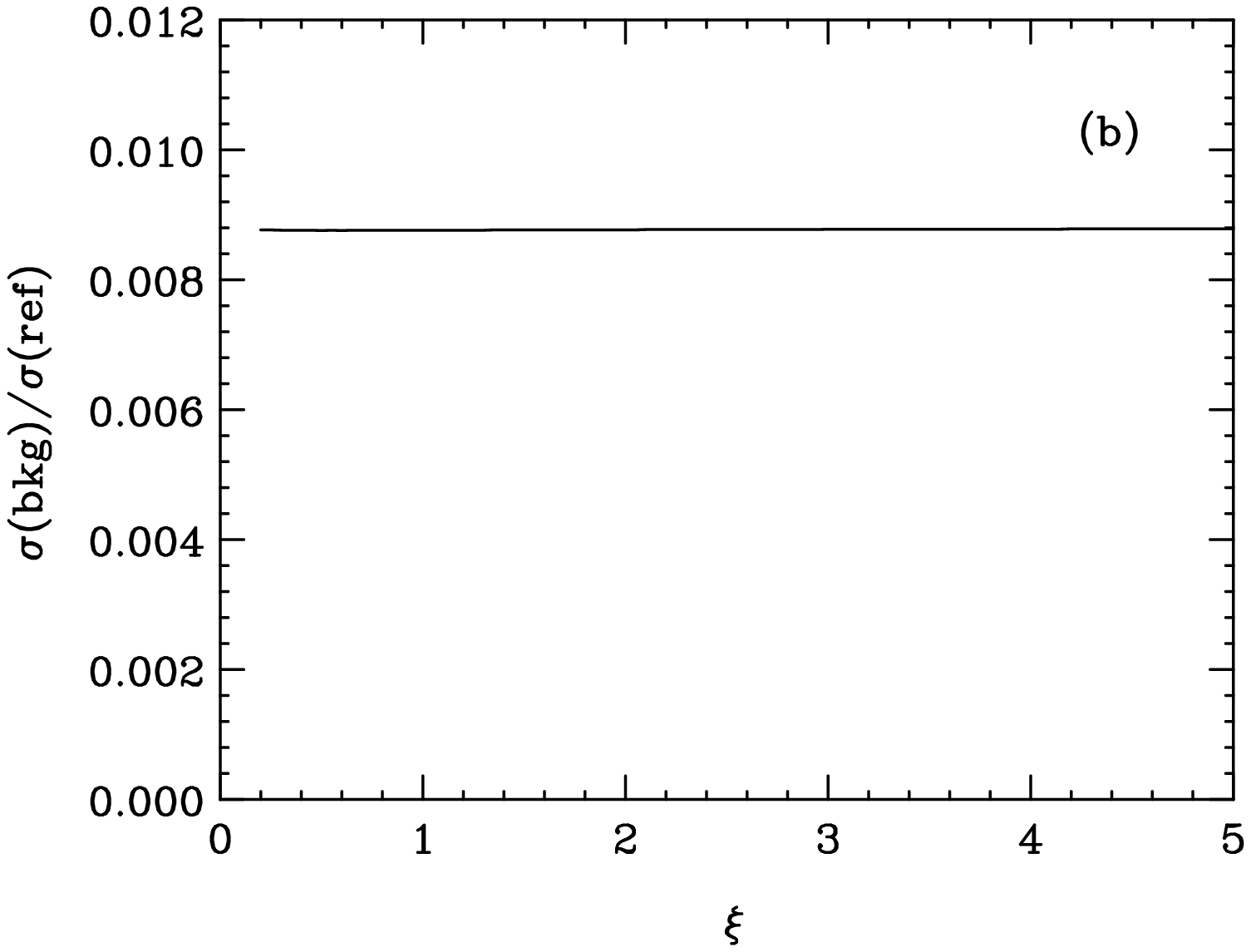} 
\end{minipage}
\caption{
  Renormalization and factorization scale variation of $t\bar{t}$ background cross
  section (a) and ratio with reference cross section (b) to 
  $H\to W^+W^-\to \ell_1^\pm \ell_2^\mp\sla{p}_T$ search in gluon fusion at the LHC
  for ATLAS GF cuts (\protect\ref{gf-atlas-cuts}, \protect\ref{gf-atlas-jet-veto})
  and $\varepsilon_{btag} = 50\%$.
}
\label{gf-scalevariation-atlas-cuts}
\end{center}
\end{figure}

\begin{figure}[htbp]
\begin{center}
\begin{minipage}[c]{.49\linewidth}
\flushright \includegraphics[width=6.cm, angle=90]{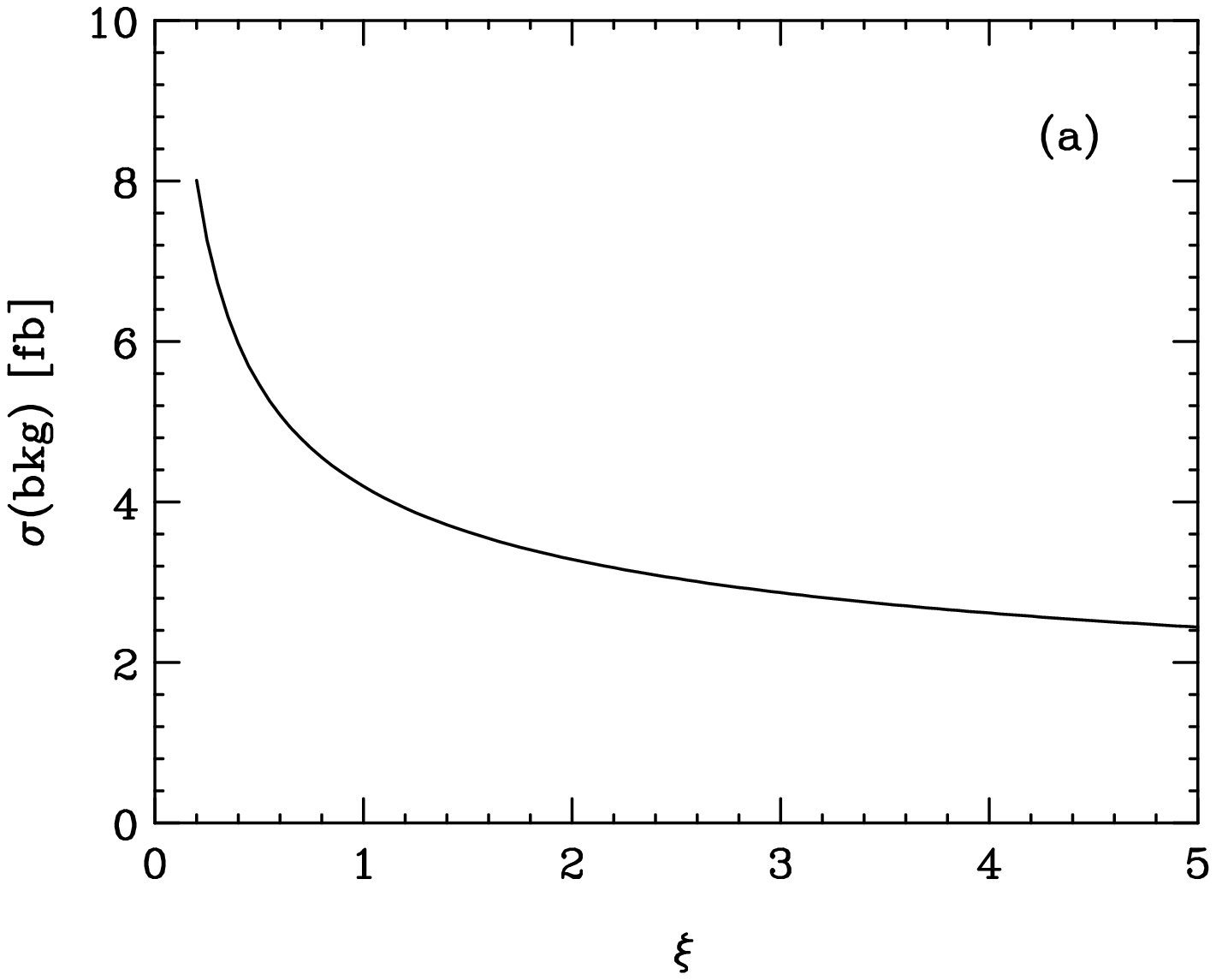}
\end{minipage} \hfill
\begin{minipage}[c]{.49\linewidth}
\flushleft \includegraphics[width=6.cm, angle=90]{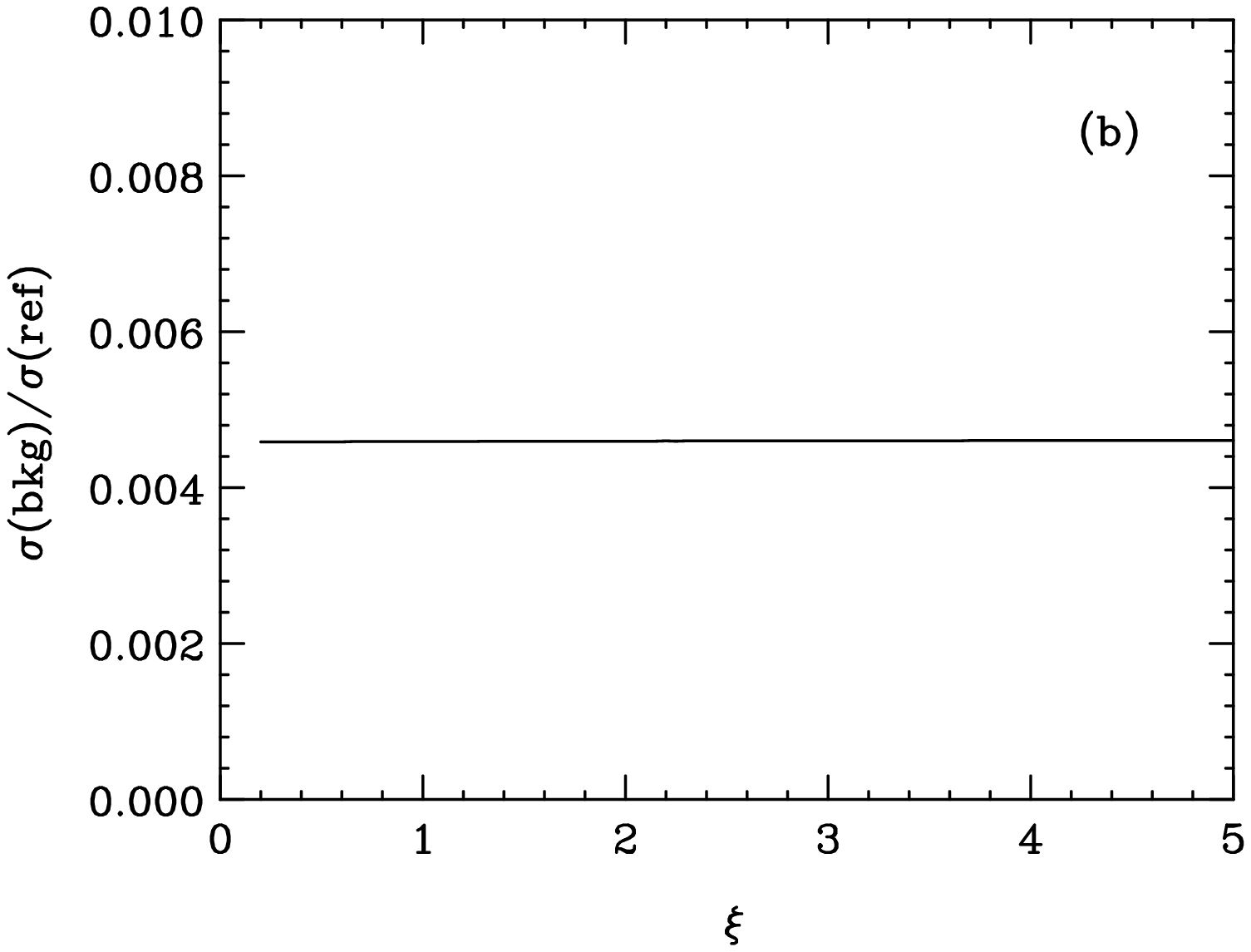} 
\end{minipage}
\caption{
 As Fig.~\protect\ref{gf-scalevariation-atlas-cuts}, but for CMS GF cuts
(\protect\ref{gf-cms-cuts},\protect\ref{gf-cms-jet-veto}).
}
\label{gf-scalevariation-cms-cuts}
\end{center}
\end{figure}

The reference cuts for the ATLAS and CMS selections are obtained by requiring at least
one tagged $b$ jet in detector region (\ref{btagcoverage}) and eliminating the central
jet veto (\ref{gf-atlas-jet-veto}) and (\ref{gf-cms-jet-veto}), respectively.
$b$ tagging capability is utilized as described in Sec.~\ref{wbf-section}.
Since the jet vetos cover most of the $b$ tagging detector region
(\ref{btagcoverage}), little additional background suppression is possible.
We therefore use $\varepsilon_{btag} = 50\%$ for all GF results.
A reference cross section of $390 \pm 97$ fb ($950 \pm 240$ fb) is obtained 
for ATLAS (CMS) selection cuts.  With 30 fb$^{-1}$ of data,
a statistical accuracy of better than 1\% can therefore be expected,
and no need to eliminate the lepton pair cuts arises for the GF selections.
Again, the scale uncertainty of the reference cross sections is very similar to that
of the background cross sections.
The scale variation of the ratio
$\sigma_{bkg}/\sigma_{ref}$ is shown in Figs.~\ref{gf-scalevariation-atlas-cuts}(b)
and \ref{gf-scalevariation-cms-cuts}(b).
It is remarkably reduced.  For the ratio, one obtains 0.0088 (0.0046) for ATLAS (CMS)
selection cuts with negligible scale variation. 

As seen in Table \ref{gf-measurement-uncertainty}, an integrated luminosity of 
30 fb$^{-1}$ would allow a measurement of the GF reference cross section
with a statistical error of better than 1\%.
Combining the uncertainty of both extrapolation factors 
yields a GF background estimate with an accuracy of better than 1\%.

\begin{table}
\caption{Expected number of events $E$ and statistical experimental error for
ATLAS and CMS GF reference selection for different integrated luminosities.
\label{gf-measurement-uncertainty}}
\vspace*{.5cm}
\begin{tabular}{|l|c|c||c|c|}
 \cline{2-5}
\multicolumn{1}{c|}{} & \multicolumn{2}{c||}{ATLAS} & \multicolumn{2}{c|}{CMS} \\
 \hline
 $\int{\cal L}\;dt$ & $E$ & $\Delta E/E$ & $E$ & $\Delta E/E$ \\
 \hline
  10 fb$^{-1}$ & 3900 & $\pm$ 1.6\% & 9500 &  $\pm$ 1.0\% \\
  30 fb$^{-1}$ & 11700 & $\pm$ 0.9\% & 28500 &  $\pm$ 0.6\% \\
  100 fb$^{-1}$ & 39000 & $\pm$ 0.5\% & 95000 &  $\pm$ 0.3\% \\
  \hline 
\end{tabular}
\end{table}

Table \ref{gf-cms-nwa-comparison} shows significant decreases of about 40--50\%
for $\sigma_{bkg}$ and $\sigma_{bkg}/\sigma_{ref}$ if matrix elements with top quark
in NWA are used instead of complete tree-level matrix elements.
This sizable decrease is caused by large sub-resonant contributions to the
jet-veto-suppressed GF backgrounds that are neglegted in NWA.  The reference cross
sections with no jet veto, on the other hand, decrease only by ca.~5\%.

\begin{table}
\caption{%
Change of background cross section and ratio for ATLAS and CMS GF selections
if calculated with complete tree-level matrix elements \protect\cite{KZ1,K2}
relative to calculating with top quark in narrow width approximation (NWA).
\label{gf-cms-nwa-comparison}}
\vspace*{.2cm}
\begin{tabular}{|c|c|c|}
 \cline{2-3}
\multicolumn{1}{c|}{} & \multicolumn{2}{c|}{$x_\text{CMS}/x_\text{NWA}$ factor} \\
\hline
$x$  & ATLAS cuts & CMS cuts \\
 \hline
 $\sigma_{bkg}$ & 2.1 & 1.7 \\
 $\sigma_{bkg}/\sigma_{ref}$ & 2.0 & 1.7 \\
 \hline
\end{tabular}\\
\vspace*{.5cm}
\end{table}

In Table \ref{gf-pdf-uncertainty} we provide an estimate for the PDF uncertainty of
backgrounds and ratios for the GF selections.
Here, the relative error is similar for backgrounds and ratios.  One obtains
about 9\% (3\%) for ATLAS (CMS) selection cuts.
The relative deviation of LO and NLO PDF results for the ratio
is about 7\% (4\%) for ATLAS (CMS) cuts.
We therefore estimate the
PDF uncertainty of the GF ratio $\sigma_{bkg}/\sigma_{ref}$ at 9\% (4\%)
for ATLAS (CMS) cuts.

\begin{table}
\caption{\label{gf-pdf-uncertainty}
GF top background cross section $\sigma :=
\sigma_{bkg}$ and cross section ratio $K:= \sigma_{bkg}/\sigma_{ref}$
calculated for ATLAS and CMS selection cuts
with PDF sets CTEQ6L1 and CTEQ6.1M (= CTEQ61.00).
The NLO sets CTEQ61.01-40 allow to calculate a PDF uncertainty for observables
(see main text, Sec.~\protect\ref{wbf-section}).}
\vspace*{.5cm}
\begin{tabular}{|l|c|c|c|c|}
\hline
\multicolumn{1}{|c|}{ATLAS} & $\sigma$ &
$\displaystyle \frac{\Delta \sigma}{\sigma}$ & $K$ &
$\displaystyle \frac{\Delta K}{K}$ \bigstrut \\
 \hline
 CTEQ6L1 (LO) & 3.3 fb & -- & 0.0088 & -- \\
 CTEQ6.1M  (NLO) & 3.2 fb & $\pm$\:8.2\% & 0.0094 & $\pm$\:9.3\% \\
  \hline 
\end{tabular}\\[.5cm]
\begin{tabular}{|l|c|c|c|c|}
\hline
\multicolumn{1}{|c|}{CMS} & $\sigma$ &
$\displaystyle \frac{\Delta \sigma}{\sigma}$ & $K$ &
$\displaystyle \frac{\Delta K}{K}$ \bigstrut \\
 \hline
 CTEQ6L1 (LO) & 4.2 fb & -- & 0.0046 & -- \\
 CTEQ6.1M  (NLO) & 3.9 fb & $\pm$\:3.0\% & 0.0048 & $\pm$\:3.1\% \\
  \hline 
\end{tabular}\\[.5cm]
\end{table}


\section{Discussion \label{disc-section}}

The approximation (\ref{extrapolationapproximation}) would become
an identity if the ratio $\sigma_{bkg}/\sigma_{ref}$ could be evaluated
to all orders in perturbation theory.  At fixed order in perturbation
theory, a scale dependence remains and, depending on the specific
scale choice, the result will deviate to a greater or lesser extent from
the exact result.\footnote{Note that this deviation is in addition to
any computational error made in the fixed order calculation.}
We refer to this error as residual theoretical error.  In practice, it
is commonly estimated from the scale variation using a prescription
like (\ref{theoryerrorestimate}).  Since differential distributions
can change significantly if new subprocesses or kinematic degrees of
freedom are activated in higher fixed order calculations, it is generally
desirable to calculate $\sigma_{bkg}/\sigma_{ref}$ and its scale variation at NLO.
A NLO analysis would also allow to obtain better estimates for the PDF uncertainties.
Unfortunately, a full NLO calculation of processes with 6 or 7 final state
particles is well beyond present capabilities.
At the time of writing a hadron collider program to calculate $t\bar{t}$ + 1 jet
production at NLO QCD with top quark in double pole approximation
is not yet available.  However, for the WBF $H\to WW$ search channel already at LO
the dominant $t\bar{t}j$ background features 3-body kinematics and quark-gluon
scattering subprocesses contribute.
One can therefore expect NLO ratios and
residual theoretical error estimates to be compatible with the ones we computed.  
On the other hand, for $t\bar{t}$ production without an additional hard
jet, i.e.~the leading top background for the GF $H\to WW$ search channel,
this is not the case.
However, for this background NLO QCD programs (in double pole approximation)
exist with full spin correlations \cite{ttinNLOwithDPAforHadrColl}
and parton shower interface \cite{mcatnlo}, which could be used in combination
with the results in Table \ref{gf-cms-nwa-comparison} to improve the
GF extrapolation analysis.  The extremely low scale variation of GF ratios
at LO suggests that the proposed reference selections will allow
the determination of the GF background with the desired accuracy of 10\% or better.

An important aspect of the full NLO correction to the studied top backgrounds
is the impact of the NLO correction to the SM top quark width, which reduces it
by about 10\% \cite{nlotopwidth}.  The top pair production cross section has
an amplified sensitivity to changes in the top quark width, as can be seen from its
LO dependence in NWA: $\sigma_\text{NWA}(t\bar{t}) \propto 1/\Gamma_t^2$.
Table \ref{disc-normaltopwidth-smalltopwidth-comparison} shows the sensitivity
of the LO top background cross section and ratio for representative WBF and
GF selections.  In both cases, the ratio is less sensitive than the
background cross section.  For the WBF channel the sensitivity is reduced
considerably.
The impact of a reduced top quark width on the off-shell matrix element
increase (Tables \ref{wbf-cms-nwa-comparison} and \ref{gf-cms-nwa-comparison})
is shown in Table \ref{disc-cms-nwa-for-smalltopwidth-comparison}.
The complete off-shell matrix element increase for the top backgrounds
is almost entirely
due to additional subresonant matrix element
contributions rather than a kinematic perturbation
of the top quark Breit-Wigner resonance distributions
through selection cuts that effectively eliminate the
central part of the distributions (instead of more
or less uniformly suppressing them).  The background
dependence on the top width is hence qualitatively similar to the
inclusive dependence, where the ratio of a single resonant to
a double resonant matrix element contribution scales approximately
linear with the width.  The complete off-shell matrix element increases
should therefore change by less than 10\% if one switches from LO to NLO
top width.  This is confirmed by the results in
Table \ref{disc-cms-nwa-for-smalltopwidth-comparison}.

\begin{table}
\caption{%
Top background cross section $\sigma_{bkg}$ and ratio $\sigma_{bkg}/\sigma_{ref}$
calculated with LO and NLO SM values for the top width $\Gamma_t$
(using $\Gamma_t(\text{NLO}) = 0.9\ \Gamma_t(\text{LO})$).
The WBF results use scale scheme (\protect\ref{transscheme}) (with $\xi = 1$)
and $\varepsilon_{btag} = 60\%$.   The GF results use ATLAS cuts (\protect\ref{gf-atlas-cuts}, \protect\ref{gf-atlas-jet-veto}).
\label{disc-normaltopwidth-smalltopwidth-comparison}}
\vspace*{.5cm}
\begin{tabular}{|l|c|c|c|c|}
\cline{2-5}
\multicolumn{1}{c|}{} & \multicolumn{2}{|c|}{WBF} & \multicolumn{2}{|c|}{GF} \\
\cline{2-5}
\multicolumn{1}{c|}{} & $\sigma_{bkg}$ &
$\displaystyle \frac{\sigma_{bkg}}{\sigma_{ref}}$ & $\sigma_{bkg}$ &
$\displaystyle \frac{\sigma_{bkg}}{\sigma_{ref}}$ \bigstrut \\
 \hline
$\Gamma_t = \Gamma_t(\text{LO})$ & 0.36 fb & 0.0032 & 3.3 fb & 0.0088 \\
$\Gamma_t = \Gamma_t(\text{NLO})$ & 0.45 fb & 0.0033 & 3.7 fb & 0.0080 \\
  \hline 
\end{tabular}\\[.5cm]
\end{table}

\begin{table}
\caption{%
As Tables \ref{wbf-cms-nwa-comparison} and \ref{gf-cms-nwa-comparison},
but cross sections calculated with NLO SM top width
$\Gamma_t = 0.9\ \Gamma_t(\text{LO})$.
Calculational details for WBF and GF selections are as in Table
\protect\ref{disc-normaltopwidth-smalltopwidth-comparison}.
\label{disc-cms-nwa-for-smalltopwidth-comparison}}
\vspace*{.2cm}
\begin{tabular}{|c|c|c|}
 \cline{2-3}
\multicolumn{1}{c|}{} & \multicolumn{2}{c|}{$x_\text{CMS}/x_\text{NWA}$ factor} \\
\hline
$x$  & WBF & GF \\
 \hline
 $\sigma_{bkg}$ & 1.15 & 1.94 \\
 $\sigma_{bkg}/\sigma_{ref}$ & 1.06 & 1.86 \\
 \hline
\end{tabular}\\
\vspace*{.5cm}
\end{table}

The WBF and GF selection cuts we applied allow collinear $g\to b\bar b$
configurations for initial state gluons, which give rise
to large log-enhanced higher-order contributions.  If these contributions
dominate, the expansion parameter of the perturbation series is
$\alpha_s\log(\mu^2/m_b^2)$ rather than $\alpha_s$, and a resummation becomes
necessary if the scale $\mu$ is of the order of $m_t$.  To detect if 
log-enhanced contributions dominate the cross sections and ratios under study, 
we calculate how much they increase if the $b$ quark mass is reduced by
a factor 100.  The results are shown in 
Table \ref{disc-normalbmass-smallbmass-comparison} and indicate that
log-enhanced contributions are small for the WBF selection cuts, but
significant for the GF selection cuts.  A resummation might thus be necessary
to obtain reliable results in the latter case.  We note that this resummation
can not be accomplished by convoluting the $b$ PDF with $gb\to bW^+W^-$,
$g\bar{b}\to \bar{b}W^+W^-$ and $b\bar{b}\to W^+W^-$ matrix elements
\cite{singletop} if
the top background is suppressed by central jet vetos like (\ref{wbf-jet-veto}),
(\ref{gf-atlas-jet-veto}) or (\ref{gf-cms-jet-veto}), or by eliminating
events with tagged $b$ jets (as described in Sec.~\ref{wbf-section}), since then
the ``spectator'' $b$ or $\bar{b}$ quark is potentially resolved and thus cannot
be integrated out to derive a suitable $b$ quark density.

\vspace*{0.cm}
\begin{table}
\caption{%
As Table \protect\ref{disc-cms-nwa-for-smalltopwidth-comparison}, but showing the
increase of the top background and ratio for WBF and GF if the $b$ quark mass
is reduced by a factor 100 (using LO SM top width and complete matrix elements).
\label{disc-normalbmass-smallbmass-comparison}}
\vspace*{.2cm}
\begin{tabular}{|c|c|c|}
 \cline{2-3}
\multicolumn{1}{c|}{} & \multicolumn{2}{c|}{$\displaystyle
  \frac{x(m_b = 0.01\:m_b(\text{SM}))}{x(m_b = m_b(\text{SM}))}$} \bigstrut\\
\hline
$x$  & WBF & GF \\
 \hline
 $\sigma_{bkg}$ & 1.2 & 2.4 \\
 $\sigma_{bkg}/\sigma_{ref}$ & 1.2 & 2.3 \\
 \hline
\end{tabular}\\
\vspace*{.5cm}
\end{table}

In addition to the discussed theoretical improvements, systematic experimental
uncertainties should also be taken into account in future studies.


\section{Conclusions \label{concl-section}}

A LO analysis was presented that demonstrates that key top backgrounds
to $H\to W^+W^-\to \ell^\pm \ell^\mp\sla{p}_T$ decays in weak boson
fusion and gluon fusion at the CERN Large Hadron Collider
can be extrapolated from experimental
data with an accuracy of order 5\% to 10\%.  A prescription to derive
the required reference selections was given.  If LO scale variation
is accepted as proxy for the theoretical error, parton level results
indicate that the $t\bar{t}j$ background to the $H\to WW$ search
in WBF can be determined with a theoretical error of about
5\%, while the $t\bar{t}$ background to the $H\to WW$ search in 
GF can be determined with a theoretical error of better
than 1\%.  Uncertainties in the parton distribution functions
contribute an estimated 3\% to 10\% to the total error.  In order to
accurately extrapolate the GF background, contributions beyond LO
should be taken into account in future studies.


\begin{acknowledgments}
We thank D.~Zeppenfeld for drawing our attention to the
extrapolation approach and N.~Akchurin, B.~Mellado and A.~Nikitenko
as well as J.~Huston, M.~Kr\"{a}mer, W.~K.~Tung and M.~Whalley
for useful comments and suggestions.  We also thank the organizers for
invitations to and hospitality at the Les Houches workshop 2003 and
the Higgs meeting during the September CMS week at CERN.    This research
was supported by the DFG Sonderforschungsbereich/Transregio 9
``Computer-gest\"{u}tzte Theoretische Teilchenphysik''.
\end{acknowledgments}


\begin{thebibliography}{99}

\bibitem{WBFcits}
D.~Rainwater and D.~Zeppenfeld,
Phys. Rev. {\bf D60}, 113004 (1999)
[Erratum-ibid. {\bf D61}, 099901 (2000)];
C.~M.~Buttar, R.~S.~Harper, and K.~Jakobs,
ATL-PHYS-2002-033 (2002);
K.~Cranmer, {\it et al.}, ATL-PHYS-2003-002 and ATL-PHYS-2003-007 (2003);
D.~Rainwater, D.~Zeppenfeld, and K.~Hagiwara,
Phys. Rev. {\bf D59}, 014037 (1999);
T.~Plehn, D.~Rainwater, and D.~Zeppenfeld,
Phys. Rev. {\bf D61}, 093005 (2000);
D.~Rainwater and D.~Zeppenfeld,
JHEP {\bf 9712}, 005 (1997) [arXiv:hep-ph/9712271];
K.~Cranmer, {\it et al.},
ATL-PHYS-2003-036 and ATL-PHYS-2003-006 (2003) [arXiv:hep-ph/0401088];
T.~Han, G.~Valencia, and S.~Willenbrock,
Phys. Rev. Lett. {\bf 69}, 3274 (1992);
T.~Figy, C.~Oleari, and D.~Zeppenfeld,
Phys. Rev. {\bf D68}, 073005 (2003);
E.~L.~Berger and J.~Campbell,
ANL-HEP-PR-04-4 (2004) [arXiv:hep-ph/0403194].

\bibitem{KPRZ}
N. Kauer, T. Plehn, D. Rainwater, and D. Zeppenfeld, Phys. Lett. {\bf B503},
  113  (2001).

\bibitem{LesHouchesHiggs2001}
D.~Cavalli, {\it et al.}, proceedings of the Workshop on Physics
at TeV Colliders, Les Houches, France, 2001 [arXiv:hep-ph/0203056].

\bibitem{Asai2003}
S.~Asai, {\it et al.},
Eur. Phys. J. {\bf C} direct, DOI: 10.1140/epjcd/s2003--01--010--8 (2003) [arXiv:hep-ph/0402254].

\bibitem{lep2higgs}
R.~Barate, {\it et al.}  [ALEPH Collaboration],
Phys. Lett. {\bf B565}, 61 (2003).
M.~W.~Grunewald, UCD-EXPH-030401 (2003) [arXiv:hep-ex/0304023].

\bibitem{GFcits}
M.~Dittmar and H.~K.~Dreiner, Phys. Rev. {\bf D55}, 167 (1997)
and [arXiv:hep-ph/9703401];
K.~Jakobs and T.~Trefzger, ATL-PHYS-2000-015 (2000);
S.~Catani, D.~de Florian, and M.~Grazzini, JHEP {\bf 0105}, 025 (2001)
[arXiv:hep-ph/0102227];
R.~V.~Harlander and W.~B.~Kilgore,
Phys. Rev. {\bf D64}, 013015 (2001) and 
Phys. Rev. Lett.  {\bf 88}, 201801 (2002);
C.~Anastasiou and K.~Melnikov,
Nucl. Phys. {\bf B646}, 220 (2002);
V.~Ravindran, J.~Smith, and W.~L.~van Neerven,
Nucl. Phys. {\bf B665}, 325 (2003);
G.~Davatz, {\it et al.}, 
CERN-PH-TH-2004-035 (2004) [arXiv:hep-ph/0402218].

\bibitem{DD2}
M.~Dittmar and H.~Dreiner, CMS-NOTE-1997-083 (1997).

\bibitem{ATLAS_TDR_2}
ATLAS Collaboration, Technical Design Report, Vol.~2, CERN-LHCC-99-15 (1999).

\bibitem{KZ1}
N.~Kauer and D.~Zeppenfeld, Phys. Rev. {\bf D65}, 014021 (2002).

\bibitem{K2}
N.~Kauer, Phys. Rev. {\bf D67}, 054013 (2003).

\bibitem{cms-scheme}
G.~Lopez Castro, J.~L.~Lucio, and J.~Pestieau,
Mod.~Phys.~Lett.~{\bf A6}, 3679 (1991);
A. Denner, S. Dittmaier, M. Roth, and D. Wackeroth, Nucl. Phys. {\bf B560},  33
(1999).

\bibitem{cteq6}
J.~Pumplin, {\it et al.}, JHEP {\bf 0207}, 012 (2002) [arXiv:hep-ph/0201195].

\bibitem{ttinNLOwithDPAforHadrColl}
W.~Beenakker, F.~A.~Berends, and A.~P.~Chapovsky,
Phys. Lett. {\bf B454}, 129 (1999);
W.~Bernreuther, A.~Brandenburg, and Z.~G.~Si,
Phys. Lett. {\bf B483}, 99 (2000);
W.~Bernreuther, A.~Brandenburg, Z.~G.~Si, and P.~Uwer,
Phys. Lett. {\bf B509}, 53 (2001);
W.~Bernreuther, A.~Brandenburg, Z.~G.~Si, and P.~Uwer,
Phys. Rev. Lett.  {\bf 87}, 242002 (2001);
W.~Bernreuther, A.~Brandenburg, Z.~G.~Si, and P.~Uwer,
CERN-PH-TH/2004-046 (2004) [arXiv:hep-ph/0403035].

\bibitem{mcatnlo}
S.~Frixione and B.~R.~Webber,
JHEP {\bf 0206}, 029 (2002)
[arXiv:hep-ph/0204244];
S.~Frixione, P.~Nason, and B.~R.~Webber,
JHEP {\bf 0308}, 007 (2003)
[arXiv:hep-ph/0305252].

\bibitem{nlotopwidth}
M.~Jezabek and J.~H.~K\"{u}hn, Nucl. Phys. {\bf B314}, 1 (1989);
A.~Czarnecki, Phys. Lett. {\bf B252}, 467 (1990);
C.~S.~Li, R.~J.~Oakes, and T.~C.~Yuan, Phys. Rev. {\bf D43}, 3759 (1991);
A.~Czarnecki and K.~Melnikov, Nucl. Phys. {\bf B544}, 520 (1999).

\bibitem{singletop}
S.~Moretti, Phys. Rev. {\bf D56}, 7427 (1997);
A.~S. Belyaev, E.~E. Boos, and L.~V. Dudko, Phys. Rev. {\bf D59},  075001 (1999);
T.~M.~P. Tait, Phys. Rev. {\bf D61},  034001  (2000);
A. Belyaev and E. Boos, Phys. Rev. {\bf D63},  034012  (2001).

\end{thebibliography}
\end{document}